\documentclass[aip]{revtex4-1}
\pdfoutput=1
\usepackage{graphicx}
\usepackage{subfigure}
\usepackage[]{figcaps}
\begin{document}
\preprint{Medical Physics}
\title{Companding technique for high dynamic range measurements using gafchromic films.}
\author{Frank Van den Heuvel}
\author{Wouter Crijns}
\author{Gilles Defraene}
\affiliation{Department of Experimental Radiotherapy,\\
Catholic University of Leuven,\\
Leuven, Belgium}
\begin{abstract}
We propose a methodology to perform dose measurements using gafchromic films which can span several decades of dose levels. The technique is based on a rescaling approach using different films irradiated at different dose levels. This is combined with a registration protocol correcting positioning and scaling factors for each film.
\par 
The methodology is validated using TLDs for out--of--field doses. Furthermore, an example is provided using the technique to characterize a small 7.5 mm sized radiosurgery cone and compared to measurements made with a pinpoint chamber.
\end{abstract}
\maketitle
\section{Introduction}
The use of gafchromic films as a measurement device has become popular during the last few years. There are many reasons for this among which:
1) high resolution,
2) no additional development needed,
3) off the shelf reading tools, and 
4) tissue equivalence.
Some disadvantages include: 1) variations in sensitivity 2) heterogeneity within a film does exist, and 3) a limited dynamic range does not allow the films to be used to quantify beam characteristics or to measure relatively low doses.  Of these disadvantages the last is the most important as batch production and pre--scanning allow to minimize the first two problems.
\par
In this paper we describe a technique to increase the dynamic range to low doses by making use of a rescaling algorithm inspired by companding noise reduction. 
Companded transmission is a technique that has been used to transmit data
over noisy telegraph lines\cite{clark1928} and has also been used in the audio--industry
to reduce cassette tape noise\cite{dolby1967an}. In essence a signal or part of a signal is amplified before transmission after which it is reduced to its normal level before being fed to the output circuit. 
\par In addition, we will show that this approach allows to perform small field measurements with superior resolution than heretofore possible.
\section{Methods and Materials}
\subsection{Films}
We use EBT--II gafchromic films provided by International Specialty
Products\footnote{ISP, New Jersey, USA}. 
Scanning the films at a resolution of 150dpi using a 10000XL Epson scanner.
The films are batch calibrated
to provide absolute dose values.
Two films represent the complete batch and is calibrated
using a two film methodology. This procedure has been described
elsewhere\cite{Crijns2011} and provided an absolute
error level of the order of 3 to 5 using the red channel output\%.
The films are calibrated to provide accurate dosimetry between 0.2Gy
and 4 Gy.
\subsection{Companding technique}
In order to provide correct data over several decades of dose (i.e. from
4 to 0.002Gy)we use different  films to and expose them to different
levels of dose.
The reference film's dose ($D_1$)is chosen in such a way that the maximal
expected dose is  close to the maximum from the dose range $[U1,L1]$
for which the film was calibrated, with $U_1$ being the upper limit
and $L_1$ the lower limit. The subsequent films are irradiated with
doses that are higher than the previous film using a multiplication
factor $F_i$. Where we denote the dose for the $i$--th film: $D_i =
F_{i-1}D_{i-1}$. The choice of this factor is made in such a way that
there is information overlap between the film. The emphasis factor ($F_i$)
used for the $i+1$--th film should be chosen so that $L_i < U_i/F_i$
\par
All films are converted to dose using a calibration protocol that links
the transmittance to dose. The specifics of this calibration procedure are
extensively dealt with elsewhere\cite{Crijns2011}. 
\subsection{Registration}
As the
films exhibit a high resolution (1 pixel $\sim$ 0.16mm) it is difficult
to exactly reproduce the position of a given film. Additionally, variations in
the beam output of a linear accelerator introduces an uncertainty in
the dose delivered and thus in the factor used to de--emphasize the
signal. Finally, variations in film sensitivity within a single batch
of films have been shown to exist and need to be taken into account.
For the previously mentioned reasons it might occur that perfect alignment
and dose levels differ from film to film. To solve this we introduced an alignment protocol taking into acount in both dosimetric and spatial variations.
This is not a straightforward problem as the geometric registration and the dose level are not independent 
parameters.  
\par
\begin{figure}[h]
\centering
\includegraphics[width=0.8\columnwidth]{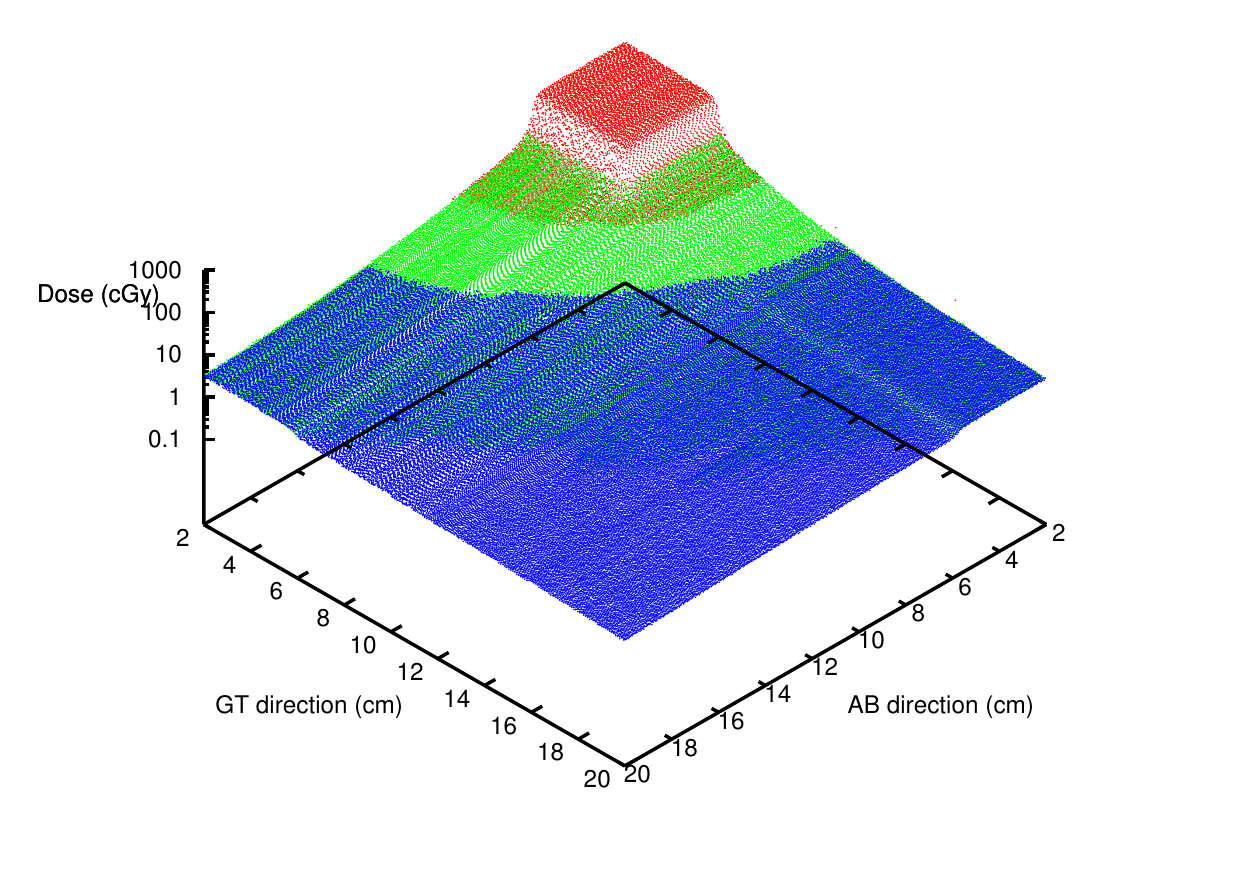}
\caption{\label{combined}Combination of three decades of irradiation
data for a 10$\times$10 cm$^2$ sized beam. The z-axis is logarithmic
and in cGy, the other axes are in cm}
\end{figure}

The registration between two films of subsequent level of companding  is performed as follows:
\begin{itemize}
\item
We define the registration as a 4--dimensional vector
$(X_i,Y_i,\theta_i,F_i)$ with $X_i$ and $Y_i$ being the translational
coordinates and $\theta_i$ a rotational parameter, the component $F_i$
denotes the emphasis factor.
\item 
The films are converted to dose using the aforementioned calibration. Only
data that falls within the upper limits ($U_i$) and lower limit ($L_i$)
of calibration is retained.
\item 
The two films are registered using data that is contained in the overlap
between the two calibration ranges. This is dependent on the position
of the film as well as the dose factor.
\item The cost function to register the images is the dose
difference between the two images multiplied by the gradient image
$G=1+a\times\sqrt{G^2_X + G^2_Y}$, with $G_X$ and $G_Y$  the gradient
calculated in the X and Y direction of the film which are chosen to
follow the film edges. The parameter $a$ is set at 500 to allow the
gradient function to have an impact on the images used as the gradient
image is normalized to have an integrated amplitude of 1.
\item The cost function is optimized using a constrained minimization
algorithm as implemented in the function {\em fmincon} in
Matlab\cite{Coleman1994Convergence}, with the option {\em interior--point approach}.
\item The minimization procedure is performed in a progressive manner
starting from sub-sampled image at 50dpi, over 75dpi and finally a
full resolution image at 150dpi. For every increase in resolution
the constraints within which the optimization is performed are
tightened. They are respectively $(5,5,10^{\mathrm{o}},10\%), (0.1,0.1,0.5^{\mathrm{o}},0.5\%)$,
and $(0.03,0.03,0.05^{\mathrm{o}},0.15\%)$. Using the same notation as for the
transformation vector with all spatial values in cm.
\end{itemize}
The end result is a dose  image of high resolution and dynamic range that
can be used to characterize a radiation beams in treatment planning
systems or to provide data on radiation dose levels far from the
field edge. An example of a resulting data set is provided in Figure
\ref{combined}.
\subsection{Validation}
The validation is performed with a single set--up. 
A stack of solid water is irradiated using a clinical 6MV photon beam
from a Varian C/D 2100 linear accelerator.
The gafchromic films are placed at a depth of 5cm, with the surface of the stack located at 95cm. We present three different tests. 
\begin{enumerate}
\item A 10$\times$10 cm$^2$ field is compared to the results from a
planning system (Varian Eclipse\texttrademark) with the pencil beam and
the AAA--algorithm. In addition the
low dose area is compared to measurements using TLDs. This data set has also been used in the Allegro project to compare with TPS data. Here we add the TLD validation.
\item A Brainlab\texttrademark$~$ radio surgery cone with an opening size of 30mm with the machine jaws set to a 5$\times$5 cm$^2$ size is compared to the data obtained for entry in the Brainscan\texttrademark$~$ planning system. For this purpose the field is measured with a PTW pinpoint chamber type 31006 of size 0.015cc\footnote{PTW Freiburg, Germany} using a waterphantom.
\item In an identical setup a cone of opening size 7.5mm was also assessed. 
\end{enumerate}
\subsubsection{TLD}
LiF:Mg,Ti TLDs (TLD--100 EXTRAD, Harshaw) are used. Calibration of the
dosimeters was per--formed in a 6 MV beam at 5cm depth in a polystyrene phantom. The TLDs are read-out in a Harshaw
6600 hot gas reader, equipped with a $^{90}$Y/$^{90}$Sr source. In order to minimize initial fading effects, read--out of calibration TLDs that have been irradiated with the internal source simultaneously with the
experiment, is part of the standard processing procedure.
\section{Results}
\subsection{$10\times10$ fields}
\begin{figure}[h!]
\centering
\includegraphics[width=0.8\columnwidth]{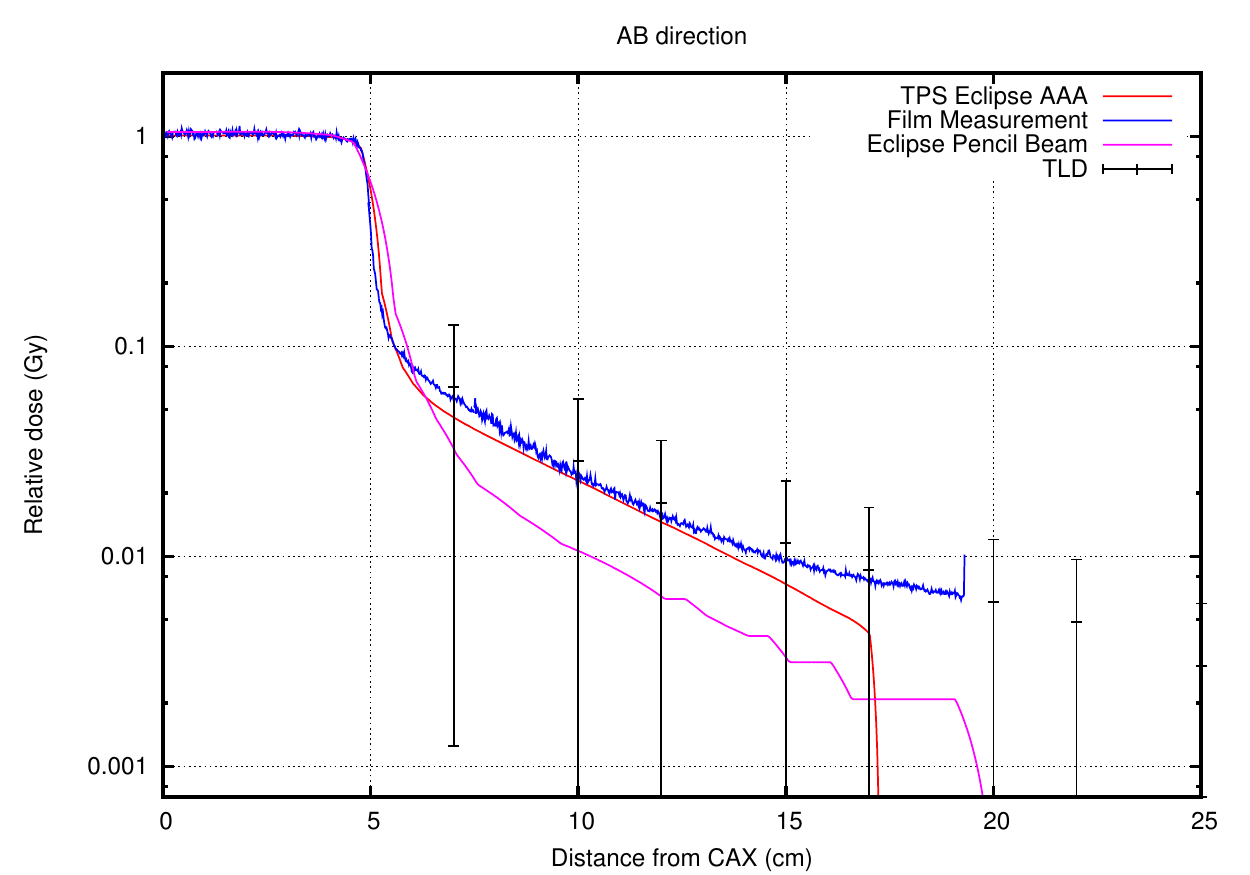}
\caption{\label{1010} Validation of a classic radiation therapy beam with a 6MV energy. The AAA implementation follows the measurement excellently. Moreover the TLD measurements show good agreement. A slight offset can be expected due to some background dose built up in the TLDs. Also note the difference in the field edge between planning and film measurement. The TPS data was modelled using data obtained with a compact IC 10 ion chamber (Wellh\"ofer).}
\end{figure}

Figure \ref{1010} provides a graph for the open field test. It is
clear that the TLD's agree well with the film measurements. Due to
the low dose on the TLD's there are considerable absolute error bars
on these measurements. Also with doses to TLDs of this magnitude the
contribution of the background can cause an offset in the measurements. A
good agreement with the planning system is found when AAA is used. The
pencil beam data does not agree that well. 

\subsection{Small field dosimetry 30mm cone}
\begin{figure}[h!]
\centering
\includegraphics[width=0.7\columnwidth]{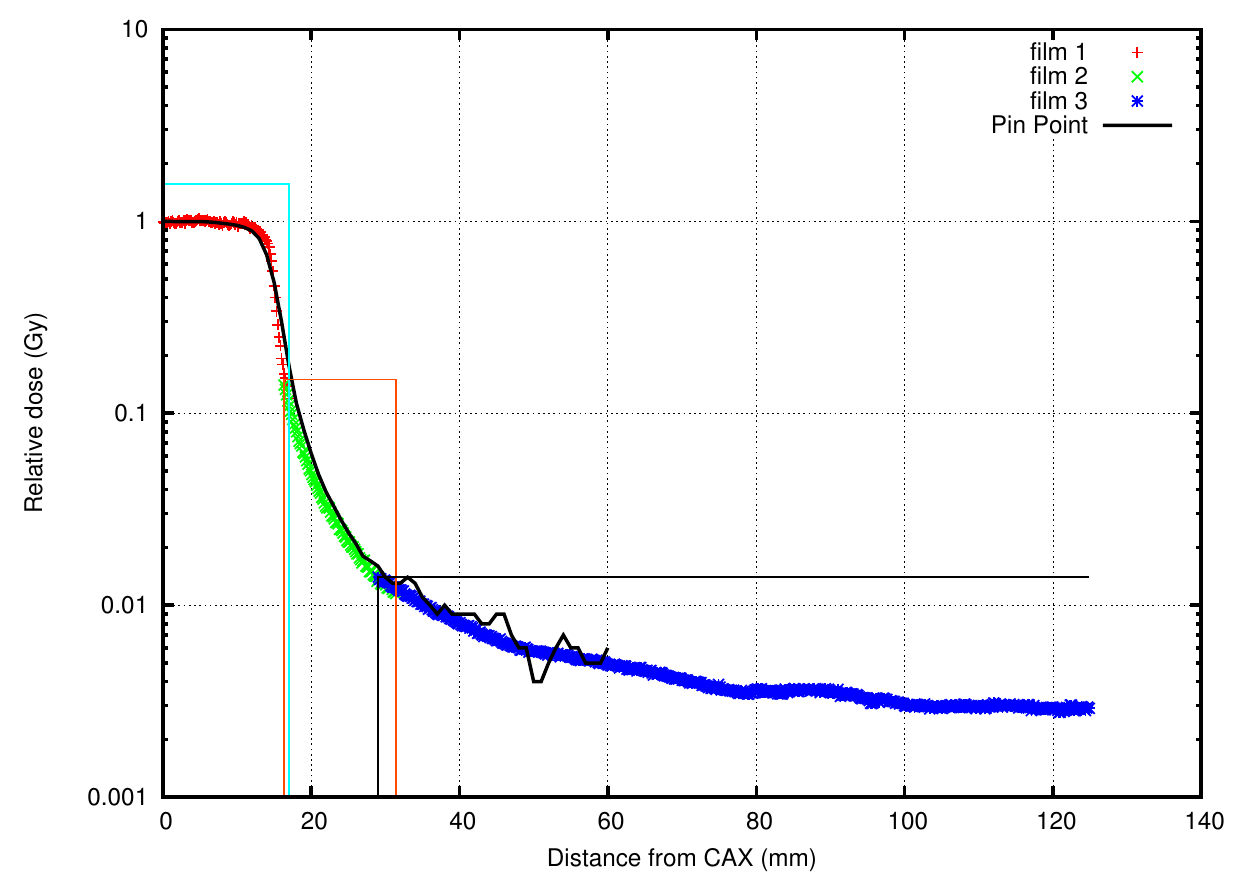}
\caption{\label{small30} A 30mm cone delivers 6MV photons on the films.We
compare to the results from a pinpoint chamber. The blue, orange and
black squares delineate the spatial and
dosimetric limitations for respectively the first, second, and third
decade. Note that even while the pin--point chamber is small a difference
can be seen in the resolving power compared to film measurements.}
\end{figure}

As in the classical field one can notice the difference in resolution
of the gafchromic film compared to the measurement. In this case a
pinpoint chamber was used. Figure \ref{small30} shows a log plot of
the relative dose. The measurement here is performed using a pinpoint chamber, however a broadening of the field edge compared to the film is easily discernible. 

\subsection{Small field dosimetry 7.5mm cone}
In the case of the data obtained with irradiation using the smallest cone there is no plateau in the data measured using the pinpoint chamber, leading us to believe that there are partial volume effects in play already. The film measurement does show a plateau phase. Also the ratio of the maximal measured dose compared to the output from a 10x10 field (output ratio) was more than double with film than with the pinpoint chamber. Due to these differences we have never used the smallest cone clinically. Furthermore, away from the central axis we note an increase in dose. Although the dose remains low increasing from 0.2\% to 0.5\% of the dose delivered to the central axis. 
\begin{figure}[h!]
\centering
\includegraphics[width=0.7\columnwidth]{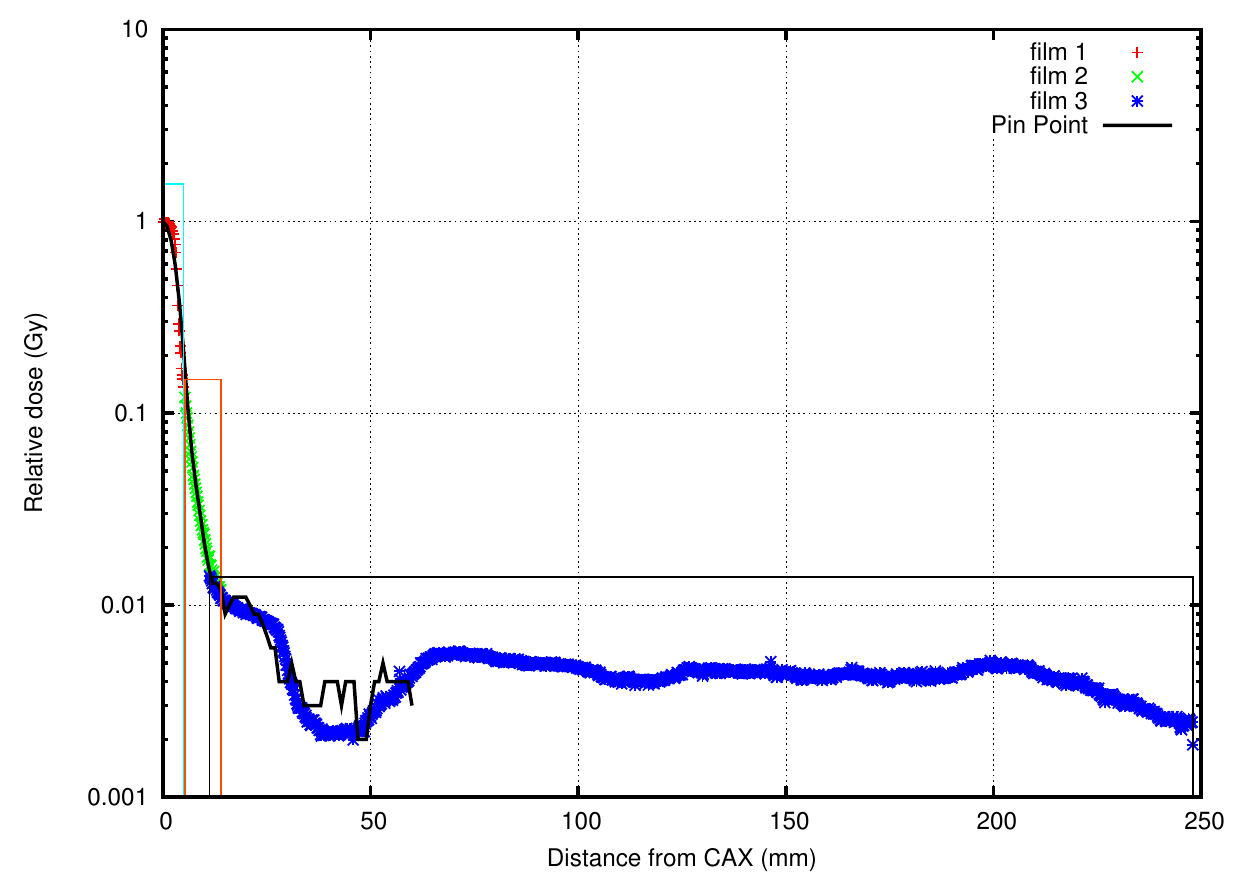}
\caption{\label{small}A 7.5mm diameter cone delivers 6MV photons on the films.We 
compare to the results from a pinpoint chamber. The blue, orange and
black squares delineate the spatial and
dosimetric limitations for respectively the first, second, and third 
decade. Note that even while the pin--point chamber is small a difference
can be seen in the resolving power compared to film measurements. Also note that further away from the central axis in the region not measured by the ion chamber the dose increases 
by an order of magnitude.}
\end{figure}
\subsection{Alignment}
In table \ref{table10} the transformation needed to match the different films show that there was a difference in response or machine output of the films used to measure the smaller dose compared to the normal film. It also shows that reasonable accuracy can be reached by mechanically positioning the films correctly in treatment as well as readout position. However, although shifts are of the order of less than 0.5 mm in the case of very steep gradients important inconsistencies can be introduced when using manual alignment only. The latter becomes important in the case of very small fields.

\begin{table}[h!]
\centering
\begin{tabular}{| l | c | c |}
\hline
{\bf 10$\times$10 Field} &Decade 1--2&Decade 1--3\\\hline
X--shift& 0.65 mm & 1.99 mm\\\hline
Y--shift& 0.64 mm & 3.35 mm\\\hline
Rotation& -0.89$\mathrm{^o}$ & -1.29$\mathrm{^o}$\\\hline
Correction for $F_i$&1.037&1.071\\\hline
\hline
{\bf Cone 30mm} &Decade 1--2&Decade 1--3\\\hline
X--shift& -0.00 mm & -0.24 mm\\\hline
Y--shift& -0.15 mm & 0.36 mm\\\hline
Rotation& -0.04$\mathrm{^o}$ & -0.33$\mathrm{^o}$\\\hline
Correction for $F_i$&0.934&0.957\\\hline
\hline
{\bf Cone 7.5mm} &Decade 1--2&Decade 1--3\\\hline
X--shift& 0.58 mm & -0.45 mm\\\hline
Y--shift& 0.07 mm & 1.53 mm\\\hline
Rotation& -2.99$\mathrm{^o}$ & -2.76$\mathrm{^o}$\\\hline
Correction for $F_i$&1.08&1.17\\\hline

\end{tabular}
\caption{\label{table10} Transformations performed to align the different films in this setup. The third film is matched to the second film. The data given are for all films with respect to the first film. This implies that errors for the third film are cumulative.}
\end{table}
\section{Discussion}
The approach as outlined here has been used to perform measurements within the framework of the ALLEGRO project. The goal of one of the workgroups was to determine the 
dose outside of the high dose volume. This to estimate the impact of newer radiotherapy techniques on the whole body dose. A second question to answer was whether data from commercial planning systems would  be able to serve as a resource to obtain information on low--dose exposures. It was therefore necessary to enhance the accuracy of low dose measurements. 
\par
From the work in this paper and the comparisons with TLD measurements and  ion--chamber data, we conclude that this technique provides data that not only allows estimates of low-dose contributions to patients, but also provides spatial information. The use of gafchromic film in this fashion will also allow its use in the characterization of radiation systems that
deliver doses to very small volumes. 
\section*{Acknowledgements}
This study was partly funded by the ALLEGRO--project (7th Euratom Framework Programme for Nuclear Research and Training, Fission--2008--3.2.1).
\section*{Author Contributions}
Mr. Van den Heuvel outlined the idea and approach and provided a proof
of concept. Mr Crijns actualized the registration component. Mr Defraene
provided the TLD  validation and the registration test. All authors
contributed to discussions and writing the article.
\bibliography{/home/fvdheuve/elsart/generic4}
\bibliographystyle{apa}

\end{document}